\documentclass{emulateapj}

\usepackage{graphicx}
\usepackage{epstopdf}

\newcommand{\wa}{{\it WISeREP}}
\newcommand{\sus}{{\it SUSPECT}}
\newcommand{\sect}{\textsection}

\shorttitle{WISeREP Supernova Repository}
\shortauthors{Yaron and Gal-Yam}

\begin{document}

\title{WISeREP - An Interactive Supernova Data Repository}

\author{Ofer~Yaron and Avishay~Gal-Yam}
\affil{Department of Particle Physics \& Astrophysics, Weizmann Institute of Science, 76100 Rehovot, Israel}

\begin{abstract}

We have entered an era of massive data sets in astronomy. In particular, the number of supernova (SN) discoveries and classifications has substantially increased over the years from few tens to thousands per year. It is no longer the case that observations of a few prototypical events encapsulate most spectroscopic information about SNe, motivating the development of modern tools to collect, archive, organize and distribute spectra in general, and SN spectra in particular.
For this reason we have  developed the {\it Weizmann Interactive Supernova data REPository} -- \wa\ -- an SQL-based database (DB) with an interactive web-based graphical interface. The system serves as an archive of high quality SN spectra, including both historical (legacy) data as well as data that is accumulated by ongoing modern programs. The archive provides information about objects, their spectra, and related meta-data. Utilizing interactive plots, we provide a graphical interface to visualize data, perform line identification of the major relevant species, determine object redshifts, classify SNe and measure expansion velocities. Guest users may view and download spectra or other data that have been placed in the public domain. Registered users may also view and download data that are proprietary to specific programs with which they are associated. 
The DB currently holds $>8000$ spectra, of which $>5000$ are public; the latter include published spectra from the Palomar Transient Factory (PTF), all of the \sus\ (SUpernova SPECTrum) archive, the Caltech-Core-Collapse Program (CCCP), the CfA SN spectra archive and published spectra from the UC Berkeley SNDB repository. It offers an efficient and convenient way to archive data and share it with colleagues, and we expect that data stored in this way will be easy to access, increasing its visibility, usefulness and scientific impact. 
We encourage the SN community world-wide to make use of the data and tools provided by \wa\ and to contribute data so that it is made globally available and archived for posterity.

\end{abstract}

\section{Introduction}

Since the realization that supernovae are a new physical phenomenon distinct from novae \citep{1938ApJ....88..529Z} and the first introduction of the spectroscopic classification of these objects \citep{1941PASP...53..224M}, the relative rarity of bright, nearby SNe that could be studied in detail precluded rapid accumulation of large spectroscopic data sets.
This state of affairs began to change in the last two decades or so, with the commencement of large, efficient supernova survey programs. These have increased the number of SN discoveries from tens to thousands per year (e.g., \citealt{2011arXiv1103.5165G}). CCD-based spectrographs with increased sensitivity now yield digital data that are easy to store and distribute.

The wealth of spectroscopic data in hand can no longer be easily described in single publications (e.g., as done in the authoritative review of \citealt{1997ARAA..35..309F}). Several groups have accumulated large amounts of spectroscopic data, some of it made publicly available (notably at the Harvard CfA, the UC Berkeley SN group, and the Asiago SN database), while specific spectroscopic databases were also established by specific projects (e.g., the Supernova Legacy Survey\footnote{http://www.cfht.hawaii.edu/SNLS/}; \citealt{2010AA...523A...7G}). In a pioneering effort the supernova group at Oklahoma University has set up \sus\ \footnote{http://suspect.nhn.ou.edu/$\sim$suspect/} \citep{2001AAS...199.8408R}, which provided a valuable community resource; but with increasing influx of data is difficult to search through and maintain. 
    
With the above in mind, we developed \wa\ (formerly known as {\it WISEASS}, standing for the Weizmann Institute of Science Experimental Astrophysics Spectroscopy System) - a new spectroscopic data repository, which is fully searchable, has a graphic interface, is professionally maintained by the Weizmann Institute of Science computing center and offers new analysis tools in addition to data archiving and distribution. 
This searchable database is already a leading repository for publicly available supernova and transient spectra, holding all of the content of the \sus\ archive, all published CCCP and PTF spectra,
the NTT/NOT SDSS-II SN spectra\footnote{http://www.physto.se/~linda/spectra/nttnot.html} \citep{2011AA...526A..28O},
the complete set of CfA Type Ia SN spectra\footnote{http://www.cfa.harvard.edu/supernova/SNarchive.html} \citep{2012AJ....143..126B} from the CfA SN archive, all published spectra from the Filippenko group at UC Berkeley (SNDB\footnote{http://hercules.berkeley.edu/database/index\_public.html}; \citealt{2012arXiv1202.2128S}) as well as additional material published by the Pan-STARRs (PS1) consortium (\citealt{2010SPIE.7733E..12K}; e.g., \citealt{2011ApJ...743..114C}) and large ESO/NTT supernova program (PI Benetti; e.g., \citealt{2011MNRAS.416.3138V}). 
\wa\ has already began accumulating and distributing spectra of the large ESO public survey - PESSTO (Valenti et al. 2012, ATel 4037), and will hold all future PTF data releases ($\sim3000$ spectra).

The aim of this report is to present the data that may be accessed and retrieved by the \wa\ DB and system, and to  introduce some of its unique features and capabilities. The paper is organized as follows:
We begin by reviewing in \sect\ref{system} the overall design and content of the DB and web-interface, followed in \sect\ref{analysis} by an explanation of the interactive online spectroscopic analysis tools. \sect\ref{access} describes how ownership of data and access permissions are handled. Specific additional utilities and cross-reference options are described in \sect\ref{utils}.
We briefly mention in \sect\ref{future} some future prospects -- existing and foreseen data sources, as well as features and additions that are being considered for future implementation;
to conclude in \sect\ref{sum} by showing exemplary sets of common (proto-type) SN spectra as well as peculiar types encountered in recent studies.

\section{System and Database Design} \label{system}

The system is based on a MySQL relational DB with a web-interface that is implemented mainly in PHP embedded within a DRUPAL -- an open-source content management system (CMS) -- framework.

\subsection{Database Tables and Relations} \label{DB}

The two major entities (implemented by the corresponding tables) are those of {\it Objects} (=sources/targets) and {\it Spectra} (with a $1:N$ relationship; i.e. several spectra may exist for each object).
Clearly, the majority of objects are SNe of the different types; however, there are also other objects for which some spectra exist in the system (e.g. novae and other cataclysmic variables; other variable stars; active galactic nuclei (AGNs); galaxies; spectrophotometric standard stars). The various object types can be reviewed in the appropriate table/web-page (see \sect\ref{web}).

The basic properties of an object are its name (whether a generally-recognized one or as attributed according to the naming convention of the relevant survey/program) and coordinates (RA,DEC). A wealth of accompanying data is specified when known; e.g. the IAU name (of a SN, if it exists); redshift and host galaxy name; discovery magnitude, filter and date; maximum magnitude, filter and date.
For both an object and a spectrum, an indication exists stating if the data is publicly available or not. Also, each object and spectrum are usually associated with a certain program (a list of the defined programs is given in the {\it Programs} web-page). A program is in many cases a name of a survey, but can also be a specific study or archive from which the data were obtained (or to which the data ``belong'').
Proprietary data are accessible only to users affiliated with the relevant programs (see \sect\ref{access}).

A spectrum is linked to an object. In addition to the ascii data file, the archive lists the following information: instrument (uniquely linked to a telescope), observation date and time, exposure time, aperture (slit), spectrum-type (object, host, sky or arcs), observer/s, reducer/s and reduction date, and optionally the corresponding FITS file (usually, the 1D wavelength+flux calibrated).
When known, additional information regarding the applied setup will be given (e.g. dichroic, grism, grating, standard star), as well as references to relevant publications. Like objects, spectra may be linked to a specific program, and marked as public or not.

Tables of the major entities (such as {\it Objects} and {\it Spectra}) include additional informative fields: remarks (free text), creation-date and last-modified (date), created-by and modified-by (username); to enable keeping track of the data creation/modification details.

For both an object and a spectrum it is possible to specify an accompanying list of free-text remarks, as well as a set of related files (additional ascii files, FITS files, graphical files or any other file types). This mode of operation supports attaching to both objects and spectra any additional required pieces of data (e.g. additional calibration products, figures and plots), without limitation on the amount of data that can be stored per entity or its nature. For example, even though the system in its current state is not formally designed to hold photometric data (but see \sect\ref{fut-tools}), it is possible to store in this way, and associate with an object, photometry-related files including images, light-curve data (tables/plots), photometric calibrations, etc.
These related-remarks and related-files are implemented by the tables: {\it Objremarks} and {\it Objfiles}, {\it Specremarks} and {\it Specfiles}, for the objects and spectra respectively.

Surrounding the main tables of {\it Objects} and {\it Spectra} (and their related entities) are auxiliary tables; these hold valuable information and may serve on their own as a useful astronomical data repository.
The major ones (which are also accessible from the main menu; see \sect\ref{web}) are: {\it Instruments}, with a $N:1$ relationship to {\it Telescopes} (both contain the basic information and links to the relevant web pages), {\it Programs}, {\it Collaborators} and a list of all the coordinates that are continuously appearing in the Astronomer's Telegrams (ATels; see \sect\ref{utils}), as well as an automatically updated list of the IAU-designated SNe.

An Entity Relationship Diagram (ERD), listing the main tables of the DB and their columns, is available online from the {\it My SQL Query} page (\sect\ref{web}).

\subsection{Web Interface} \label{web}

\begin{figure}
\plotone{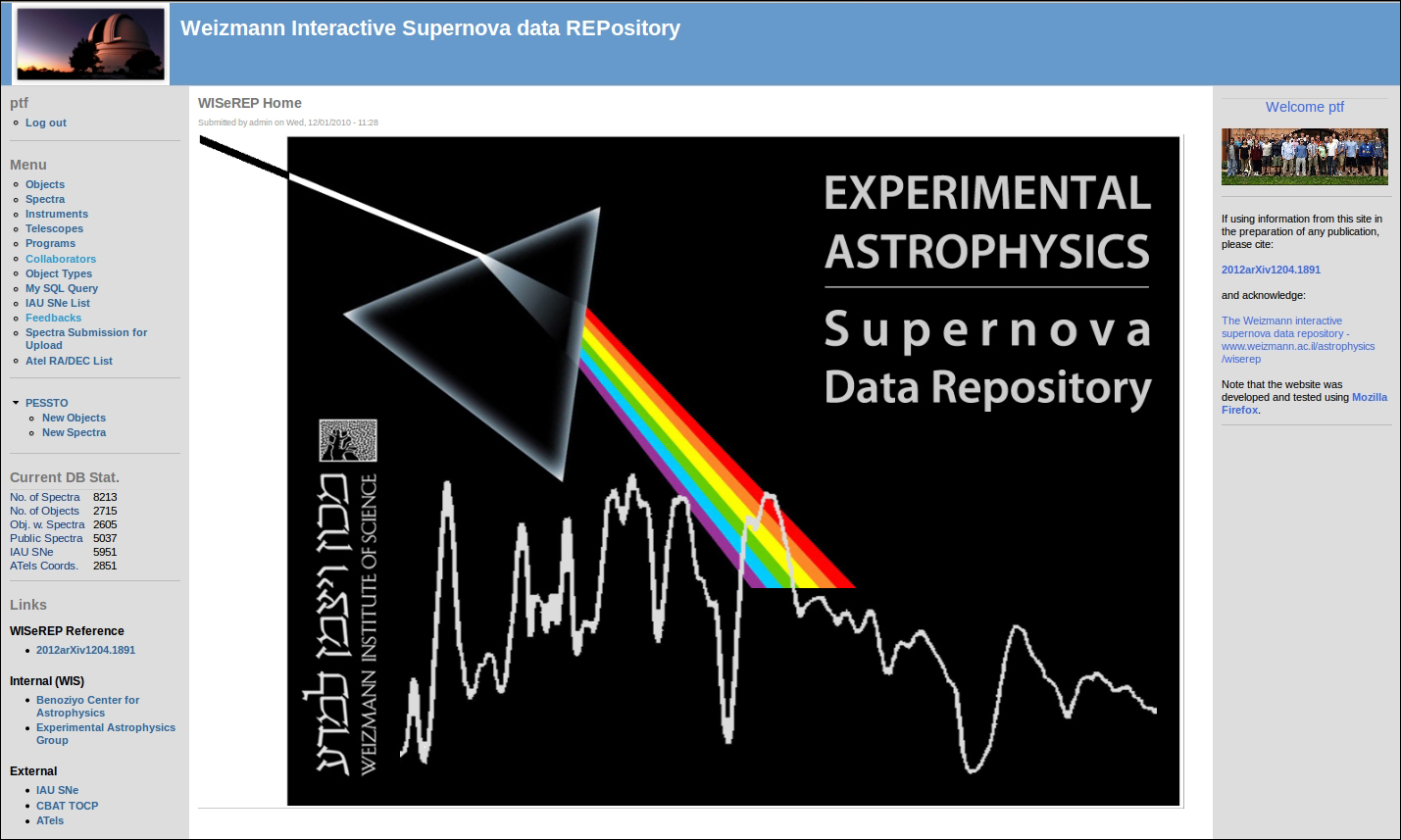}
\caption{A screenshot of the {\it homepage} after login (of user {\it ptf}  in this example). Every user account is associated with a certain {\it role}, according to which access permissions are set. The main menu is seen on the left. Besides various auxiliary tables (such as telescopes, instruments, programs, object types), the main entities are those of {\it Objects} and {\it Spectra} (the first two on the menu). The required spectra/data may be reached via either of those two pages. Instructions for submitting spectra for upload to the archive appear on the page: Spectra Submission for Upload.\label{fig-homepage}}
\end{figure}

Upon login at URL: http://www.weizmann.ac.il/astrophysics/wiserep, with either public/anonymous or a provided personal or group account, a main menu appears on the left (Fig.~\ref{fig-homepage} displays a snapshot of the homepage). Statistical information about the content of the DB and some selected links, internal and external (of e.g. major SN archives and surveys), are shown below the main menu.

Data can be accessed via the two pages dedicated to {\it Objects} and {\it Spectra}.
Generally, each web page consists of a query fields section (including sort-by options); upon submission of the query, the resulting rows are displayed below.

{\it Objects} page - Objects may be retrieved according to the following query fields: object name (either by selecting from a drop-down list of all existing names or by typing free-text to include any part of the name), the type of object (via selection from list or free-text) and program name. It is also possible to query by coordinates, either RA/DEC between certain ranges or a cone search around a specific RA-DEC, and also according to the general informative details of when the object was created or last modified (date ranges) and by whom (free-text username). Sort-by options include the name, object type, RA and the program name.
The details of the retrieved objects include the majority of the {\it Objects} table fields as listed above, with links to a variety of cross-references (both external and internal tools and data repositories) according to the object's coordinates (see \sect\ref{utils}).
If spectra, related files, or related remarks exist for the object, these can be shown by clicking the appropriate links in the object's entry.

{\it Spectra} page - Query fields similar to the above are available here, with the addition of: spectrum type (object, host, sky, arcs), instrument/telescope, ranges of dates for observation and reduction dates, observer/s and reducer/s. It is also possible to query by any part of a publication/contribution specification. Sort-by options include, among others, e.g. object name, type, observation date, observer and instrument.
The details of the retrieved objects include the majority of the {\it Spectra} table fields as listed above, with links for downloading the spectrum ascii/FITS files and any existing related files. If the number of retrieved spectra exceeds a certain value (currently set to $50$) then only static plots of the spectra are shown; otherwise, interactive plots are displayed, which enable zooming, binning and identification of major SN/galactic lines (\sect\ref{analysis}).

\begin{figure}
\plotone{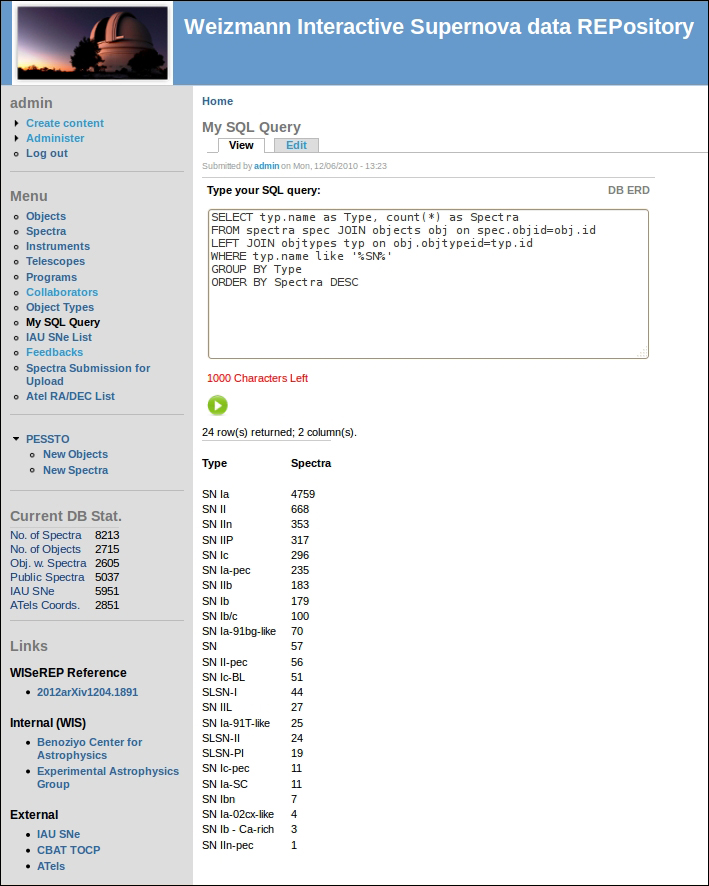}
\caption{A screenshot of the {\it My SQL Query} page. Registered users having basic knowledge in writing SQL SELECT statements may compose their own queries. Shown here is an example of a query displaying the number of spectra that currently exist in the database of each SN type. An Entity Relationship Diagram, listing the tables of the database and their columns, is provided to help compose queries.\label{fig-mysql}}
\end{figure}

{\it My SQL Query} page - Registered users that are familiar with the DB structure and SQL syntax may construct their own queries. A link for displaying the Entity Relationship Diagram (ERD), which lists the main tables of the DB and their columns, is provided to help compose such queries. See Fig.~\ref{fig-mysql} for an example.

{\it Spectra Submission for Upload} page - the current server configuration of the system - website+DB - is such that it is possible to insert/update/upload data only from within the Weizmann institute. Outside users may view and download data (according to their access permissions) but cannot modify it. Therefore, a procedure has to be followed for submitting spectra and data for upload to the system; the instructions are described on this web page.
The procedure basically involves creating a tar-ball of the ascii files (and optionally FITS files and additional related files), preparing a list of the accompanying meta-data (by filling an existing template), and sending these two items to the archive administrator.

It should be emphasized that it is possible to upload spectra that should not be made available to the public (e.g. unpublished material). Such spectra (and their objects) can be linked to a certain {\it program}, and only those accounts that are members of this program (see \sect\ref{access}) may obtain access to the data.
Users can thus use \wa\ to archive and share proprietary data, and make use of the analysis tools, without committing to immediate public release of their data.
At any given stage, objects and/or spectra can be updated as "public" (i.e. available to the public) or "non-public", following the instructions of the data owners.

Additional entries exist in the menu for some auxiliary entities, such as {\it Instruments}, {\it Telescopes}, {\it Programs} and {\it Object Types}. These are also valuable data repositories (that may serve the occasional astronomer); for instance, for querying and reviewing the information of many instruments/telescopes, which includes direct links to their specific web sites.

The {\it Atel List} page will be described in \sect\ref{utils}.

\section{Online Basic Analysis Tools} \label{analysis}

A novel feature of the \wa\ archive is that it provides online analysis tools that can be interactively applied to selected data.
Interactive plots, with zooming and over-plotting capabilities, can be used to visualize the data and inspect its quality and utility to the user. 
The plots also provide a graphical interface for interactive line identification, useful to determine host redshifts, to investigate supernova 
elemental composition and classification, and to determine ejecta expansion velocities.
Whenever a redshift is known and specified for the object, the plots display both the observed and the rest-frame wavelength scales. A list of major elements (with specific ionization levels) that are commonly identified in SNe of different types are shown next to each plot of a spectrum.
When a checkbox of a chemical element is marked, its lines are over-plotted and it is possible to specify and modify both the redshift (by default, the object's redshift, if specified) and an expansion velocity, which leads to the lines being redshifted or blueshifted, respectively. Hovering with the mouse over a checkbox displays the lines (wavelengths) of the corresponding element that are currently defined for over-plotting. Besides the pre-defined set of elements, it is possible to specify custom wavelengths for plotting, and there is also an option to over-plot common galaxy lines (which include the Balmer series, NII, OII, OIII, NaI, MgII, SII and CaII H\&K). This option is useful for determining the redshift, when it is not yet known, using narrow emission/absorption features (of the host galaxy) that may be visible in the spectrum. Atmospheric A and B absorption bands (telluric lines) can also be displayed, which can serve for both identifying artifacts that may remain from telluric line removal, or for providing a sanity check on the wavelength solution of a spectrum.
Zooming on specific regions is possible using mouse selection, and different spectral binning may be applied for visualization.

\begin{figure}
\epsscale{1.0}
\plotone{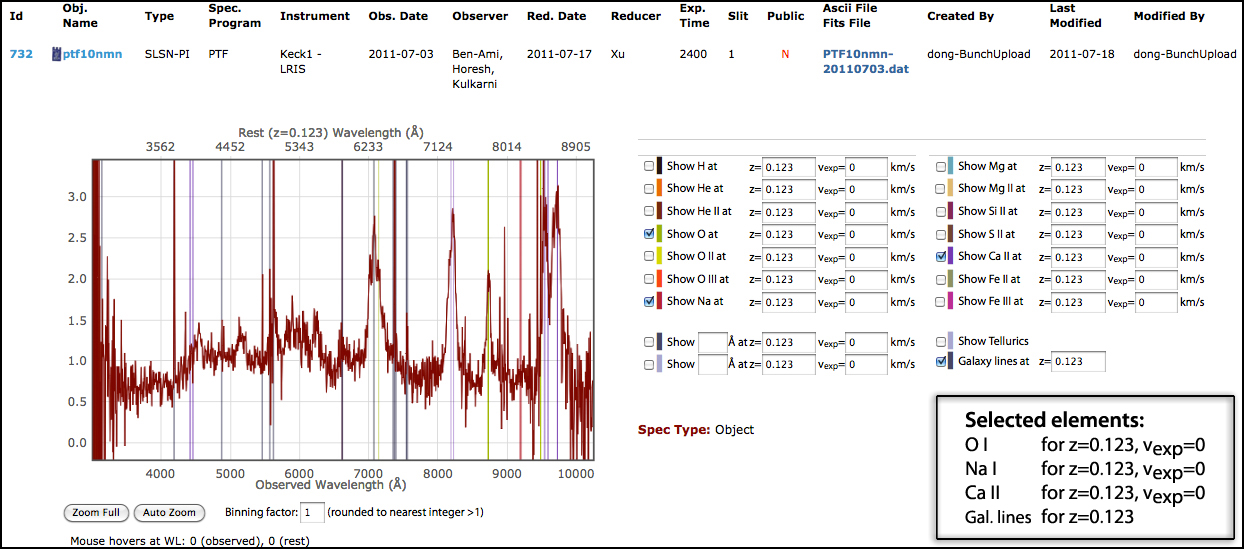}
\caption{A section of a screenshot of the Spectra page, showing an example of line identifications in the Keck1-LRIS nebular spectrum of the PISN candidate - PTF10nmn. The plot is interactive and enables live zooming, binning and over-plotting. The top x-axis marks the rest-frame wavelengths whenever the redshift is known and indicated. Note here the over-plotting of a selection of O I, O II, Na I and Ca II lines for the appropriate redshift and for zero expansion velocity, corresponding to an emission line nebular spectrum.
Common galaxy lines are also over-plotted for the given redshift, aligning with the narrow host features (which can serve for both checking the validity of the spectrum wavelength solution, or for determining the redshift when it is not already known).
Inset at the lower right points out the selected elements and parameters.
\label{fig-10nmn}}
\end{figure}

\begin{figure}
\plotone{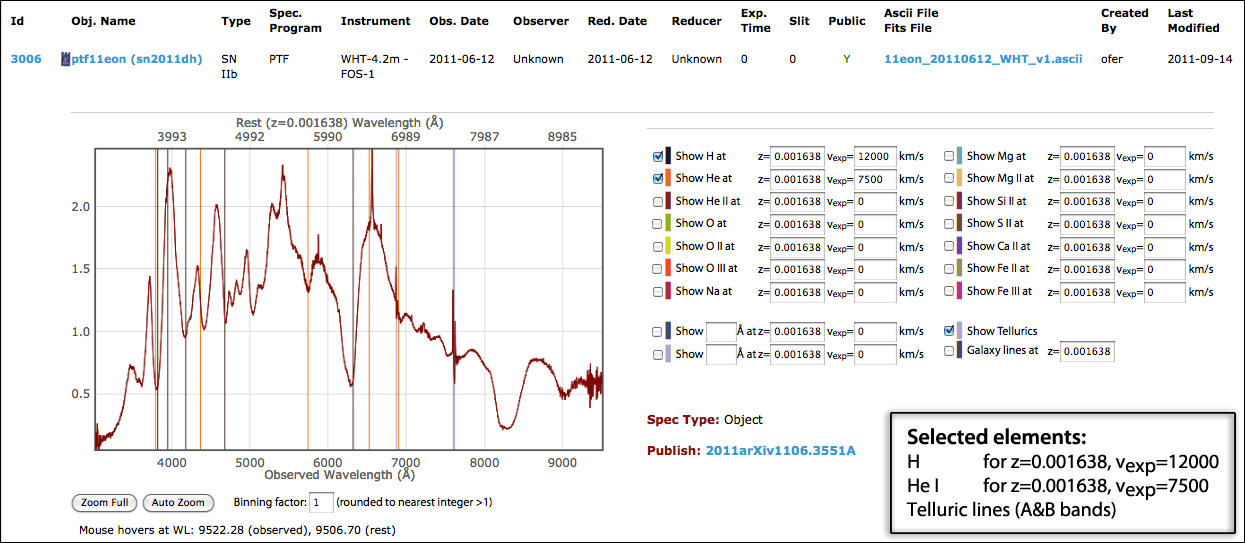}
\plotone{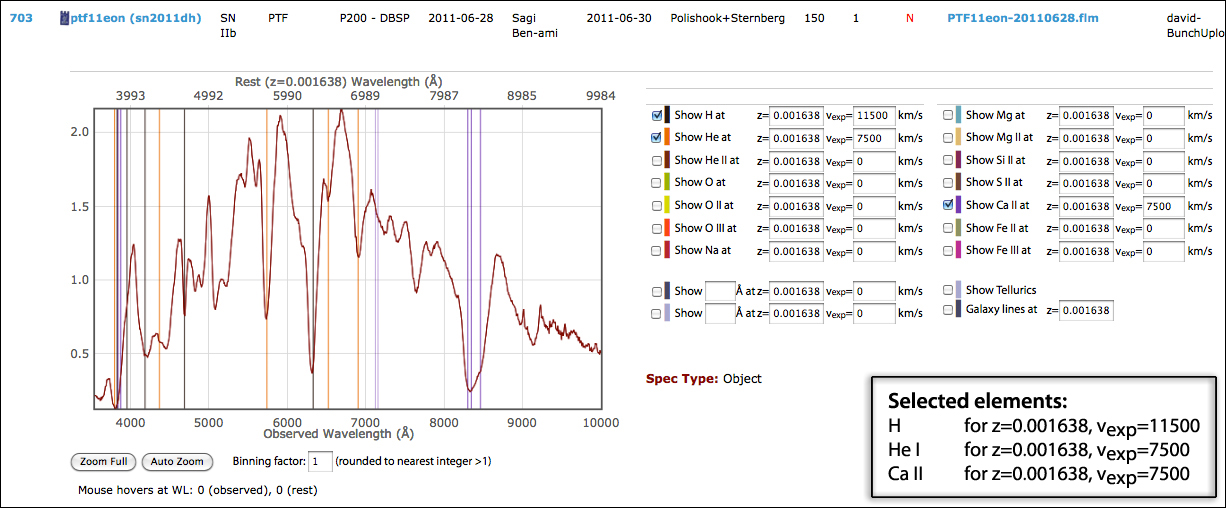}
\caption{An example of line identification for two spectra of the recent Type IIb SN in M51 - SN2011dh (PTF11eon; \citealt{2011ApJ...742L..18A}). Top: WHT spectrum from June 12.  Lines of H (Balmer series) and He I are over-plotted for the appropriate redshift and expansion velocities - $12000$ and $7500$ km s$^{-1}$, respectively. Note that whereas He lines are clearly emerging, the $\lambda 6678$ line is obscured here by the narrow $H\alpha$ emission line from the underlying host galaxy. Also plotted for demonstration are the telluric lines, which explain the corresponding artifacts seen on the spectrum. Bottom: P200-DBSP spectrum from $\sim2$ weeks later. Lines of H, He~I and Ca~II are plotted for expansion velocities of $11500$ and $7500$ km s$^{-1}$.
Insets at the lower right corners point out the selected elements and parameters.
\label{fig-11eon}}
\end{figure}

Examples of line identifications for two different SNe are shown in Figures.~\ref{fig-10nmn} and \ref{fig-11eon}.
In Fig.~\ref{fig-10nmn} we display the identification of major emission lines in a nebular spectrum of a candidate pair-instability
SN (PISN) from the PTF survey - PTF10nmn (Yaron et al. 2012, in preparation); with a redshift of $z=0.123$, zero expansion velocities are applied as appropriate for an emission line nebular spectrum.
In Fig.~\ref{fig-11eon} we display examples of line identifications in two spectra, separated by $\sim2$ weeks, of the recent Type IIb SN in M51 - SN 2011dh (PTF11eon; \citealt{2011ApJ...742L..18A}). See captions for details.

The interactive plots are useful to perform quick and efficient analysis online, determine SN types, redshifts and expansion velocities, and to select specific data for further detailed studies offline.

\section{Accounts, Ownership of Data and Access Privileges} \label{access}

The purpose of \wa\ is to serve as a data repository for communities world-wide; for holding, sharing and analyzing spectra and data that are either public or private to a certain collaboration/program. The system therefore supports a modular management of ownership and access privileges.

A user can either log into the system as an occasional guest, in which case only the public data can be accessed;
or, being a member of a specific collaboration/survey/program that is storing data in the archive, can log in using a personal or group account that has been provided. 
Each registered user account can be associated (via a mediating {\it role} definition) with one or several specific {\it programs};
whereas, from the other side, both objects and spectra can also be associated with a certain program. 
In this manner all access permissions and ownership of data are managed and followed.

It should be emphasized that every object and every spectrum can be independently linked to a program.
Thus, for example, an object can be linked to a certain program, but it may have spectra that belong to other programs.
Of those spectra, some could be marked as public (available to all), but those that are not publicly available can only be viewed and accessed by the accounts that are members of the corresponding programs.

This handling of ownership and access privileges enables each group/collaboration to not only use the archive to distribute 
data to the community, but also to share data that remains proprietary to the group members. Groups can thus use the archive as a
useful collaboration platform, as well as use the analysis capabilities, without committing to a public data release. 
At any point in time, every object or spectrum can be individually marked as public or non-public. Users can choose to keep non-public data totally private
(such that no other users know that these objects or spectra even exist), or to allow the system to show other users that additional data exist in the
system, but are proprietary to specific groups; this mode will motivate collaborative efforts.

\section{Additional Utilities and Cross-References} \label{utils}

As part of the \wa\ website we maintain additional useful utilities that are not necessarily an integral part of the archive, but offer convenient querying possibilities that may also serve those who are not necessarily searching for specific SN spectra or data.

The {\it ATEL RA/DEC List} page enables searching Astronomer's Telegrams according to coordinates; between values of RA/DEC, or a cone search around a given RA/DEC.
The possibility to search by coordinates is not provided by the ATEL website\footnote{http://www.astronomerstelegram.org/} (there is no specific format for providing RA/DEC values), and is clearly useful in various situations; e.g. for cross-referencing a new transient/object in hand, or to check (according to its coordinates) if it has been recently announced or not.

An entry exists in our DB for every coordinate that is specified in an ATEL, and within those entries it is also possible to query by ATEL information: post date, title, keywords and authors.
Upon submission of a query, clicking an ATEL number in the retrieved list brings up the telegram's page from the ATEL website.

The list, a result of parsing each newly-added ATEL entry, is automatically updated every hour and contains the majority of object RA/DEC coordinates reported in ATELs (missing $<\sim5\%$ in total and complete to all new entries, including identification of PNV/PSN/TCP -- standing for possible nova/possible SN/other variable type -- coordinates).

From both the {\it Objects} page and the {\it ATEL RA/DEC List} page there are links (according to the object coordinates) to the following cross-references: 
(1) ADS\footnote{http://adsabs.harvard.edu/} (The SAO/NASA Astrophysics Data System) - position search results;
(2) NED\footnote{http://ned.ipac.caltech.edu/} (NASA/IPAC Extragalactic Database);
(3) SDSS (Sloan Digital Sky Survey) Navigator Tool\footnote{http://cas.sdss.org/dr5/en/tools/chart/navi.asp} (currently DR8);
(4) SIMBAD\footnote{http://simbad.u-strasbg.fr/simbad/} astronomical database;
(5) MAST\footnote{http://archive.stsci.edu/} (The Multimission Archive at STScI) DSS FITS image retrieval;
(6) WISE\footnote{http://irsa.ipac.caltech.edu/Missions/wise.html} (Wide-field Infrared Survey Explorer) All-sky Release image data via the NASA/IPAC IRSA (Infrared Science Archive);
(7) PTF image and source archive at the NASA/IPAC IRSA;
and (8) NASA's HEASARC\footnote{http://heasarc.gsfc.nasa.gov/} (High Energy Astrophysics Science Archive Research Center).

From the {\it Objects} page there is also an internal link to the {\it ATEL RA/DEC List} page to automatically retrieve corresponding ATEL entries (if these exist) using the object coordinates.

\section{Future Prospects} \label{future}

\subsection{Sources - SN Surveys and Archives} \label{fut-surveys}

The archive currently holds SN spectroscopic data from several sources and programs.
These include the complete \sus\ archive ($>1500$ spectra for $\sim180$ SNe, all public);
the $2603$ Type Ia SN spectra obtained during $1993-2008$ through the Center for Astrophysics (CfA) Supernova Program (\citealt{2012AJ....143..126B});
published spectra from the UC Berkeley SNDB ($\sim330$ public spectra for $\sim90$ SNe; \citealt{2012arXiv1202.2128S});
the NTT/NOT SDSS-II SN spectra ($\sim240$ public spectra for $\sim190$ SNe; \citealt{2011AA...526A..28O});
PTF (\citealt{2009PASP..121.1334R}, \citealt{2009PASP..121.1395L}; $\sim3000$ spectra in total, of which $\sim60$ are published spectra made public);
CCCP -- the Caltech Core-Collapse Project \citep{2007ApJ...656..372G, 2012ApJ...744...10K} -- currently including $\sim180$ spectra of which $\sim20$ are public;
HST-Ia (observations of SNe Ia with HST (cycles 17; 18; PI Ellis); $75$ spectra, internal to program members);
HIRES (a study of Type Ia SN using high-resolution spectroscopy - \citealt{2011Sci...333..856S}; $60$ public spectra);
PS1 public spectra (\citealt{2011ApJ...743..114C});
public spectra from the large ESO/NTT program \citep{2011MNRAS.416.3138V, 2011MNRAS.412.2735T}, and several other sources.
We expect that additional data will be ingested into \wa\ in the near future from existing archives.

\wa\ will also serve as the curation and distribution facility for several ongoing and future surveys. In particular,
the \wa\ system has began to serve as the distribution platform of the ESO large public spectroscopic PESSTO survey, 
which will host and distribute extensive public datasets of SN spectra and associated meta-data and information on targets.
PESSTO (acronym for: {\it Public ESO Spectroscopic Survey for Transient Objects}; see Valenti et al. 2012 - ATel 4037) is a $4(+1)$ year survey planned to deliver detailed, high-quality time series optical+NIR spectroscopy of $\gtrsim100$ optical transients; to be obtained with the ESO-NTT telescope located in La Silla Paranal Observatory - Chile.
\wa\ will also serve as the data portal of future PTF data releases (over $3000$ spectra already archived).

\subsection{Additional Features and Tools} \label{fut-tools}

A major addition that is being considered for future implementation is extending the functionality of the system to include the archiving and analysis of photometric data; thus forming a more complete information reservoir of high quality SN data.
In particular, a graphic interface and automatic analysis of photometry (e.g. determination of peak SN dates, light curve fitters) is foreseen.

An extension of spectral analysis capabilities is also planned, to include, e.g.:
line fitting routines, to determine line widths and equivalent widths;
online application of global spectral fitting software (e.g., the SuperNova IDentification tool {\it SNID} - \citealt{2007ApJ...666.1024B}, and the supernova spectral fitting package {\it Superfit} - \citealt{2005ApJ...634.1190H});
queue-based application of parametric spectral synthesis routines ({\it SYNAPPS} - \citealt{2011PASP..123..237T})
and extension of spectral tools to include the IR and UV ranges.

High level fitting results may be archived and distributed to further increase the utility of the data.

\section{Exemplary SN Spectra and Summary} \label{sum}

\begin{figure}
\plotone{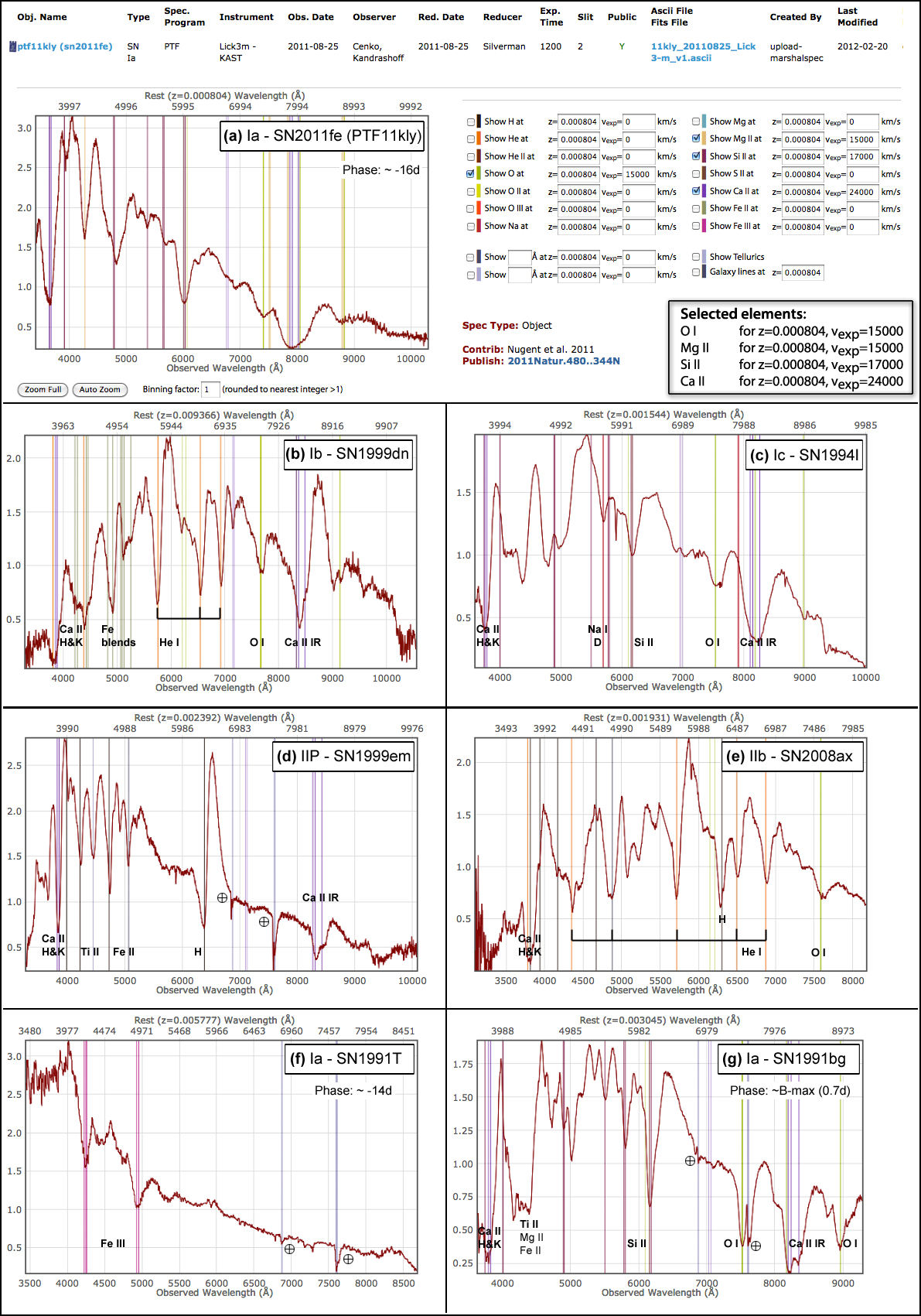}
\caption{Plots of representative early-time spectra from the \wa\ archive (that originate from various programs - e.g. \sus, PTF), denoting  major SN types. Panel a - for the recent Type Ia SN 2011fe - displays a complete snapshot of the spectrum entry - the interactive plot alongside all related information that appears in the top (header) row and the element check-boxes area to the right. The remaining panels show only the graphs with the element identification lines over-plotted; major features are marked. See Table \ref{tab-sntypes} for details of the shown spectra.
\label{fig-specs}}
\end{figure}

\begin{figure}
\plotone{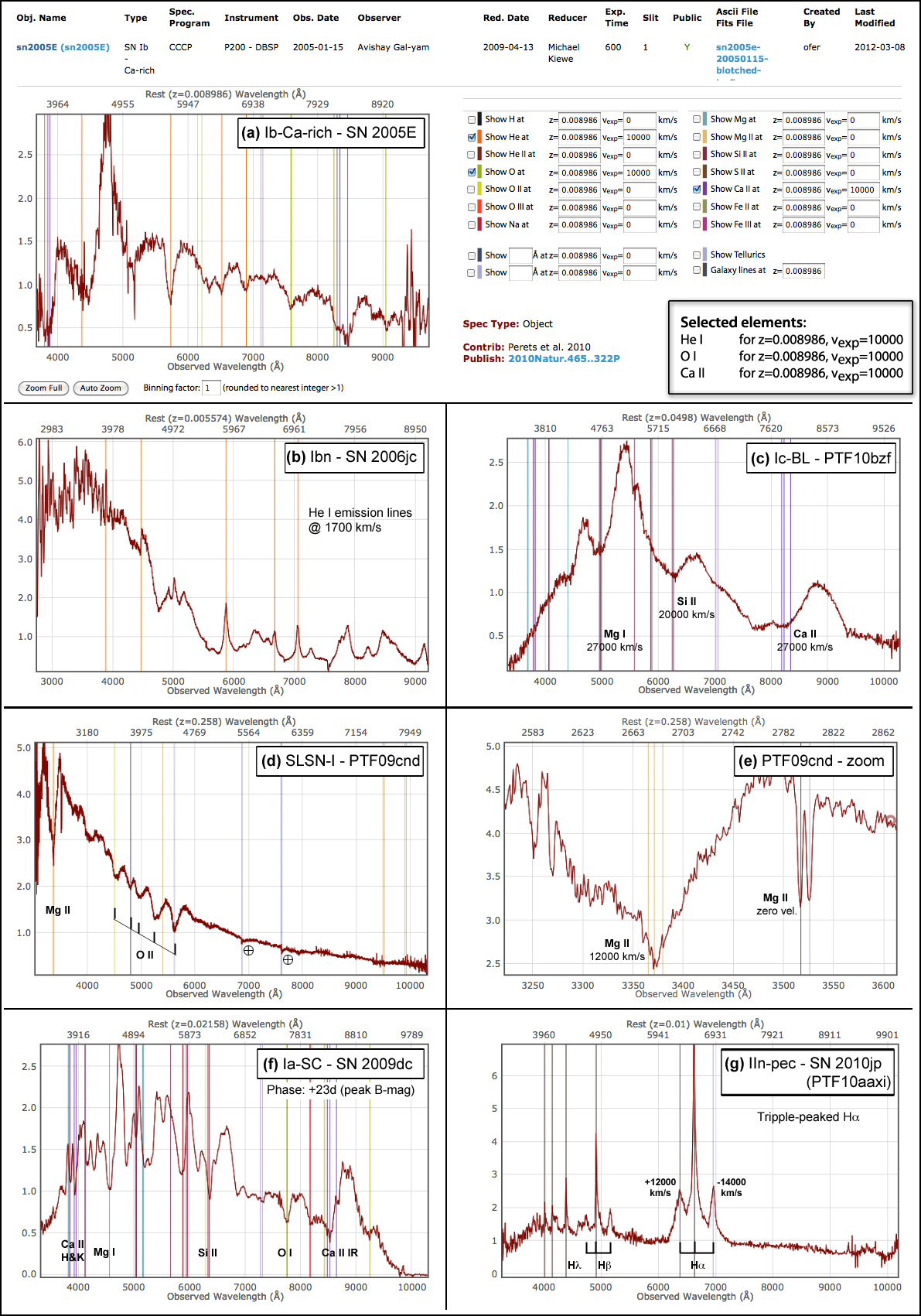}
\caption{A representative set of spectra from the \wa\ archive, denoting some of the more {\it peculiar/exotic} SN types.
Like Fig.~\ref{fig-specs}, panel a - for the Type Ib Ca-rich SN 2005E - displays a complete snapshot of the spectrum entry - interactive plot, related information and meta-data. The other panels show only the graphs with the element identification lines over-plotted. 
Panel e is a blow-up of the Mg II features region (broad blue-shifted absorption and narrow doublet host absorption) of the PTF09cnd spectrum shown in panel d. See Table \ref{tab-pectypes} for details.
\label{fig-pecspecs}}
\end{figure}

To further deliver a sense of the content and appearance of the system, as well as to provide a useful primer 
to the classification of common supernova spectra, 
we show in Figures \ref{fig-specs} \& \ref{fig-pecspecs} two representative sets of spectra.
Fig.~\ref{fig-specs} displays a set of seven spectra obtained around peak light ($\pm2$ weeks) of common SN types, whereas in Fig.~\ref{fig-pecspecs} we show a set of spectra that serve as examples for some of the more peculiar cases we have encountered in recent years. See Tables \ref{tab-sntypes} and \ref{tab-pectypes} for the corresponding details.

To conclude, we have reported here on the development of a new spectroscopic system that is fully searchable and offers online analysis tools in addition to data archiving and distribution. The interactive plots are useful to perform quick and efficient analysis online, determine SN types, redshifts and expansion velocities.

The archive already hosts $>8000$ spectra (of which $>5000$ are public) for $\sim2600$ objects, including the complete \sus\ and {\it CfA-Ia} samples, and will serve as the curation and distribution facility for several ongoing and future surveys (e.g., the PESSTO survey and PTF data releases).

At any point in time, objects and spectra can be marked as publicly available or not.
Only the users that are members of a relevant program may view and download its proprietary data.
Therefore, users and groups can use \wa\ as a useful collaboration platform; to archive and share proprietary data, and make use of the analysis tools, without committing to immediate public release of their data.

We encourage the SN community world-wide to make use of the data and tools provided by \wa\ and to contribute data so that it is archived for posterity and made easily available to the community, thereby increasing its utility and the visibility of the paper reporting it.
For sending data for upload to the system, please follow the instructions specified in the {\it Spectra Submission for Upload} page.

We are aware that there may still exist erroneous data (or meta-data), missing information or duplications of spectra (within a specific program or across programs), especially within the bulk of spectra ingested from the \sus\ archive (e.g., ascii files of spectra that have been uploaded with the wavelengths de-redshifted and/or flux de-reddened, instead of the non-manipulated reduced spectra). 
We will appreciate being notified whenever such cases are encountered so that we are able to correct and improve the content of this data repository continuously and on a regular basis.

The URL address of \wa~is: \url{http://www.weizmann.ac.il/astrophysics/wiserep}.
It is possible to log in as a guest user; members of programs supported by \wa, please send a request for a personal (or a group) account by following the instructions in the homepage.

For technical support or any questions that may arise, please contact \email{wiserep@weizmann.ac.il}.
General comments, feedbacks and suggestions are clearly most welcome.

\acknowledgments

We would like to thank the \sus\ team -- Eddie Baron, Mary Hogan and Jerod Parrent  -- for their help and cooperation in ingestion of the complete \sus\ archive into \wa. In this respect, many thanks also to Raphael Schepps and Adam Becker, for their devoted help and effort in organizing and cleaning of \sus\ data.
We are grateful to the CfA SN team, in particular Stephane Blondin and Bob Kirshner, for their kind help and cooperation in ingestion of the CfA spectral dataset into \wa.
Special thanks to Derek Fox, Brian Schmidt and Stephen Smartt for their helpful involvement and concern in various discussions;
and to Jeff Silverman for his continuous cooperation and valuable contribution (and specifically, with regards to ingestion of the UC Berkeley SNDB data).
We highly appreciate the assistance of Giorgos Leloudas and Ariel Goobar of Stockholm university, for their help with the incorporation of the complete SDSS-II NTT/NOT sample.
We thank Iair Arcavi for his contribution to the development of the interactive plots.



\bibliographystyle{apj}
\bibliography{mybibfile}



\clearpage

\begin{table}
  \caption{Details of the spectra of {\it common (proto-type)} SN types shown in Fig.~\ref{fig-specs}}
  \begin{tabular}{l l l l l}\label{tab-sntypes}
    \\
    Type & SN Name & Obs. Date & Telescope-Instrument & Reference \\
       	 & \multicolumn{4}{l}{Remarks} \\
    \hline \hline
    Ia	& PTF11kly		& 2011-08-25	& Lick3m - KAST	& \cite{2011Natur.480..344N}	\\
        	& \multicolumn{4}{l}{Ions typical of pre-max Type Ia (O I, Mg II, Si II, S II, Ca II, Fe II).} \\
    \hline
    Ib	& SN 1999dn	& 1999-09-17	& Lick3m - KAST	& \cite{2001AJ....121.1648M}  \\
        	& \multicolumn{4}{l}{He I, Ca II, Fe II @ $9000$ km s$^{-1}$.} \\
    \hline
    Ic	& SN 1994I		& 1994-04-09		& Ekar-1.82m - B\&C	& \cite{1996ApJ...462..462C}	 \\
        	& \multicolumn{4}{l}{Phase: $-1.5$d (before peak light). O I, Na I, Si II @ $10000$ km s$^{-1}$; Ca II @ $15000$ km s$^{-1}$.} \\
    \hline
    IIP	& SN 1999em	& 1999-11-14	& ESO-NTT - EMMI	& \cite{2001ApJ...558..615H} \\
        	& \multicolumn{4}{l}{$\sim2$ weeks post peak. H, Ca II @ $9000$ km s$^{-1}$;} \\
        	& \multicolumn{4}{l}{Fe II $\lambda5185$, Ba II/Ti II $\lambda4552$ @ $7500$ km s$^{-1}$.} \\
    \hline
    IIb	& SN 2008ax	& 2008-03-31	& P200 - DBSP	& \cite{2008MNRAS.389..955P} \\
        	& \multicolumn{4}{l}{A few days post peak. H @ $13000$ km s$^{-1}$; He I (incl. $\lambda5016$) @ $9000$ km s$^{-1}$.} \\
    \hline
    Ia-91T	& SN 1991T	& 1991-04-16	& ESO spectrum	& \cite{1995AA...297..509M}	 \\
        	& \multicolumn{4}{l}{Clear absence of Si II, Ca II at early times. Dominant Fe III lines @ $14000$ km s$^{-1}$.} \\
    \hline
    Ia-91bg	& SN 1991bg	& 1991-12-14	& Ekar-1.82m - B\&C	& \cite{1996MNRAS.283....1T}	 \\
        	& \multicolumn{4}{l}{O I, Si II, Ca II @ $10000-12000$ km s$^{-1}$. Blends of Ti II, Mg II, Fe II at $\lambda$'s $4200-4500$.} \\
    \hline
  \end{tabular}
\end{table}

\begin{table}
  \caption{Details of the spectra of the more {\it peculiar (exotic)} SN types shown in Fig.~\ref{fig-pecspecs}}
  \begin{tabular}{l l l l l}\label{tab-pectypes}
    \\
    Type & SN Name & Obs. Date & Telescope-Instrument & Reference \\
       	 & \multicolumn{4}{l}{Remarks} \\
    \hline \hline
    Ib-Ca-rich	& SN 2005E		& 2005-01-15	& P200 - DBSP	& \cite{2010Natur.465..322P}	\\
        	& \multicolumn{4}{l}{He I, O I, Ca II @ $10000$ km s$^{-1}$.} \\
    \hline
    Ibn	& SN 2006jc	& 2006-10-23	& UVOT - UV-grism	& \cite{2009ApJ...700.1456B}  \\
        	& \multicolumn{4}{l}{He emission lines @ $1700$ km s$^{-1}$.} \\
    \hline
    Ic-BL	& PTF10bzf	& 2010-03-07	& Keck1 - LRIS	& \cite{2011ApJ...741...76C}	 \\
        	& \multicolumn{4}{l}{(SN 2010ah); Very high velocities $\sim30000$ km s$^{-1}$.} \\
    \hline
    SLSN-I	& PTF09cnd	& 2009-08-25	& Keck1 - LRIS		& \cite{2011Natur.474..487Q} \\
        	& \multicolumn{4}{l}{A set of O II lines at velocities $\sim 12000$ km s$^{-1}$, and Mg II @ $13000$ km s$^{-1}$.} \\
    \hline
    Ia-SC	& SN 2009dc	& 2009-05-18	& ESO-NTT - EFOSC2	& \cite{2011MNRAS.412.2735T}	 \\
        	& \multicolumn{4}{l}{Intermediate mass elements @ $7000$ km s$^{-1}$.} \\
    \hline
    IIn-pec	& SN 2010jp	& 2010-12-30	& Keck1 - LRIS	& \cite{2012MNRAS.420.1135S}	 \\
        	& \multicolumn{4}{l}{(PTF10aaxi) $+75$d from discovery. Tri-peaked Balmer series.} \\
    \hline
  \end{tabular}
\end{table}

\clearpage

\end{document}